\documentstyle[aps,prl,psfig]{revtex}
\begin{document}
\draft
\title{Magnetic Roughness and Domain Correlations in 
Antiferromagnetically Coupled Multilayers}
\author{Sean Langridge and J\"org Schmalian}
\address{Rutherford Appleton Laboratory, Chilton,
Didcot, Oxfordshire, OX11 0QX, United Kingdom.}
\author{C.H. Marrows, D.T. Dekadjevi and B.J. Hickey}
\address{Department of Physics and Astronomy, E.C. Stoner Laboratory,
University of Leeds, Leeds, LS2 9JT, United Kingdom.}
\date{\today}
\maketitle
\begin{abstract}
The in-plane correlation lengths and magnetic disorder of magnetic
domains in a transition metal multilayer have been studied using
neutron scattering techniques. A new theoretical framework is
presented connecting the observed scattering to the in-plane
correlation length and the dispersion of
the local magnetization vector about the mean macroscopic
direction. The results unambiguously show the highly correlated
nature of the antiferromagnetically coupled domain structure
vertically throughout the multilayer. We are
easily able to relate the neutron determined magnetic dispersion
and domain correlations to magnetization and magnetotransport
experiments.
\end{abstract}

The interplanar coupling and in-plane magnetic domains are
essential to an understanding of the origin of the large giant
magnetoresistance effect (GMR) \cite{baibich} in magnetically
coupled multilayers. This coupled with the advances in thin film
deposition techniques \cite{smith} has led to a huge interest in
magnetic multilayer systems specifically with respect to their
device application possibilities. The GMR effect arises from the
antiferromagnetic (AF) coupling of typically a transition metal
ferromagnet (e.g. Co) across a noble metal non-magnetic spacer
(e.g. Cu). This AF coupling can be realized by tuning the noble
metal spacer thickness \cite{parkin}. The change in resistivity
results from the spin dependent scattering of the conduction
electrons which depends not only on the magnetic moment alignment
but also on the interfacial disorder \cite{zahn} and the magnetic
domain structure. Until recently the question of the relationship
between magnetic domain structure and interlayer coupling has not
been explored experimentally. It is clear that a vertically
incoherent magnetic domain structure will have the effect of
lowering the GMR by preventing perfect AF alignment in adjacent
layers \cite{holloway}. In studies of a weakly coupled system such
as [Cu(60\AA)/Co(60\AA)]$\times$20 it was shown that the reduction
in the GMR from the as-prepared state to the coercive state can be
understood as a loss of vertical coherence of the AF coupling
\cite{borchers}. The situation is different in the strongly
coupled samples investigated here, the results of which clearly
show magnetically correlated domains at the coercive field which
extend vertically throughout the entire multilayer.

The investigation of {\em structurally} rough interfaces is well
established and makes use of diffuse x-ray scattering techniques.
The theoretical tools for analyzing various surface morphologies
are well advanced \cite{sinha1,sinha2,holy,pynn}. Recent advances
in x-ray techniques have applied this structural formalism to the
study of magnetically rough systems
\cite{chen,freeland,idzerda,mackay,kao,fischer,chakrian}.
Nevertheless, the problem of quantifying magnetic roughness
remains difficult primarily due to the indirect and complicated
nature of the spin-photon interaction\cite{gibbs,hannon}. This
problem can be resolved by neutron techniques for which the direct
interaction between the neutron's dipole moment and the sample
magnetization is well understood.

In this letter we have performed neutron scattering measurements
on magnetically coupled multilayers and quantitatively determined
the field dependence of the magnetic roughness and domain
distribution. The large lateral coherence length of the neutron
beam ($>30\mu$m \cite{webster}) ensures that the measurements
sample many magnetic domains. Since the neutrons are highly
penetrative the measurements also sample the whole multilayer
vertically, unlike the transition metal L$_{\rm III}$ x-ray
measurements \cite{hase} which sample primarily the uppermost
interfaces because of the high x-ray absorption coefficient.

We prepared Co/Cu and Co/Ru multilayers of 50 bilayer repeats,
with Cu and Ru spacer thicknesses corresponding to the 1st and 2nd
AF maxima of the coupling oscillation, for different thicknesses
of the magnetic layer. The samples were deposited by dc magnetron
sputtering in a custom vacuum system with a base pressure of
2$\times$10$^{-8}$ Torr. The multilayers were grown on 20
mm$\times$25 mm pieces of (001) Si wafer with the native oxide
layer left intact. The working gas was 3 mTorr of Ar, and
deposition rates for Co, Cu and Ru were all $\sim$3 \AA/s. Smaller
10 mm$\times$2 mm samples were grown in the same growth run for
magnetoresistance and Magneto-Optic Kerr Effect (MOKE)
measurements.

\begin{figure}
\psfig{figure=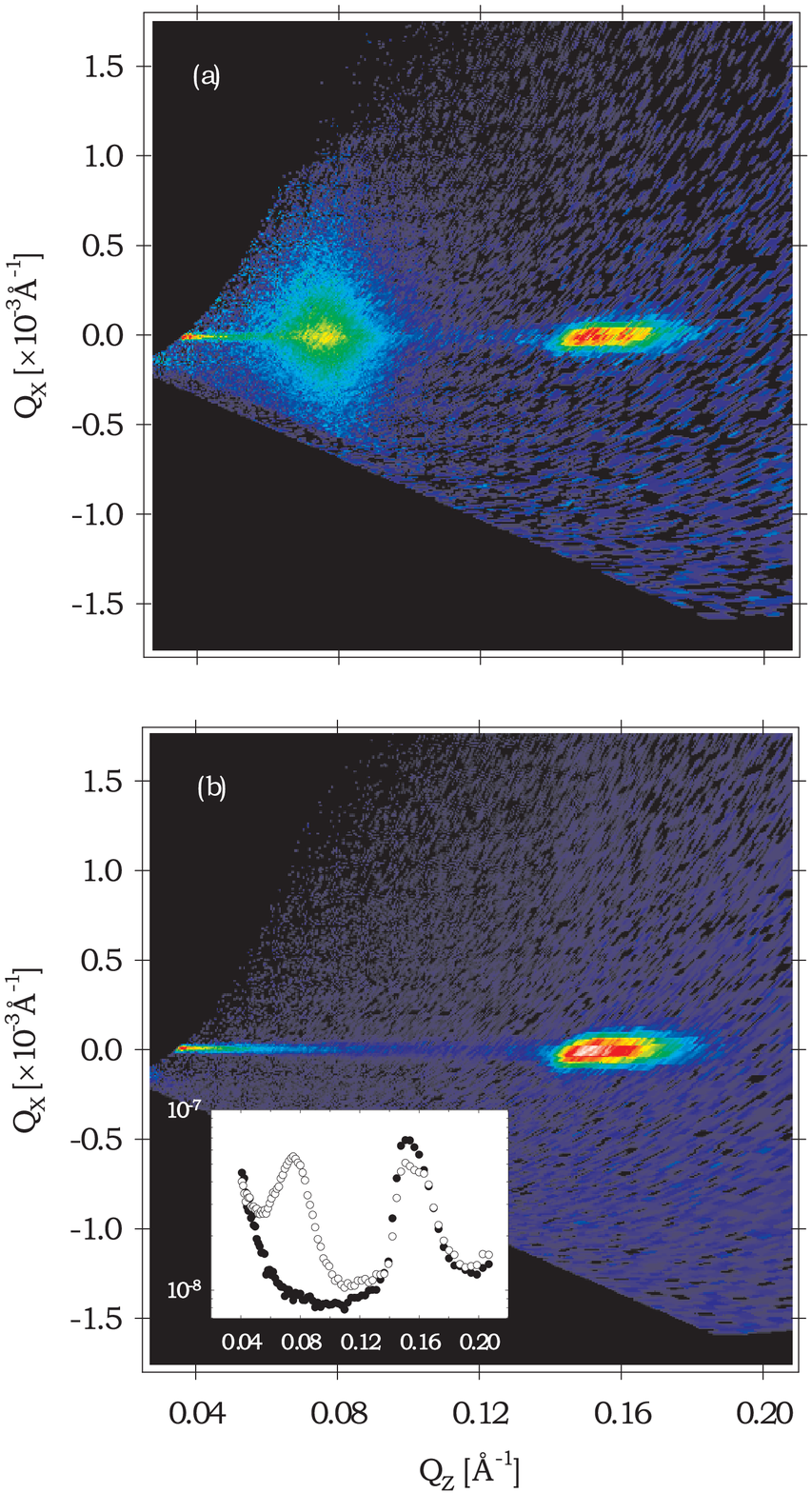,width=8.5cm} \caption{(a): The
observed scattering from the [Co(20\AA)/Cu(20\AA)]$\times$50
multilayer in zero applied field. The intensity centered at
$Q_Z$=0.075 \AA$^{-1}$ corresponds to the AF ordering wave-vector
and arises purely from the magnetic ordering. The intensity at
twice this wave-vector is the first order multilayer structural
Bragg peak. The dark areas represent the kinematical limits of the
measurement. (b) The corresponding measurement in a saturation
field of H=700 Oe. The AF correlations are suppressed leaving only
the specular ridge ($Q_X$=0) and the first order Bragg peak. The
inset shows the specular reflectivity for the low (open symbol)
and high (closed symbol) field data.} \label{frsm}
\end{figure}

The reflectivity measurements, both polarized and non-polarized,
were performed on the time-of-flight polarized neutron beam
reflectometer CRISP at the ISIS facility, Rutherford Appleton
Laboratory \cite{felici,crisp}. To maximize the flux at the sample
position, for the diffuse scattering measurements, the
reflectometer was run in a non-polarized mode with an incident
wavelength range of 0.5 \AA-6.5 \AA. An electromagnet at the
sample position provides an in-plane reversible field of $\pm$7
kOe. The scattered neutrons are detected by a 1-dimensional $^3$He
detector. The combination of the time-of-flight technique and the
multidetector ensure that both the parallel ($Q_Z$) and
perpendicular ($Q_X$) (to the surface normal) components of the
neutron wave-vector transfer (see fig \ref{frsm}) are obtained in
a single measurement. Typical acquisition times are of the order
of 2 hours for an entire reciprocal space map, which compares
favorably with resonant x-ray techniques \cite{hase}.

Fig.\ref{frsm}(a) presents the observed reciprocal space
intensity map for the nominal [Co(20\AA)/Cu(20\AA)]$\times$50
multilayer at remanence. This Cu thickness corresponds to the 2nd
AF ordering peak. Although we have similar data for other Co, Cu
and Ru layer thicknesses we shall concentrate on this sample in
this article. Three features are apparent in the data: the
specularly reflected ridge ($Q_X$=0 \AA$^{-1}$), the first order
nuclear Bragg peak ($Q_Z$=0.15 \AA$^{-1}$) and the Bragg peak
corresponding to the AF periodicity ($Q_Z$=0.075 \AA$^{-1}$). This
peak is entirely magnetic in origin. The first order Bragg peak
indicates that the bilayer thickness is $\sim$42 \AA. The narrow
width in $Q_Z$ (see inset) implies that the AF order is coherent
throughout the whole multilayer.

\begin{figure}
\psfig{figure=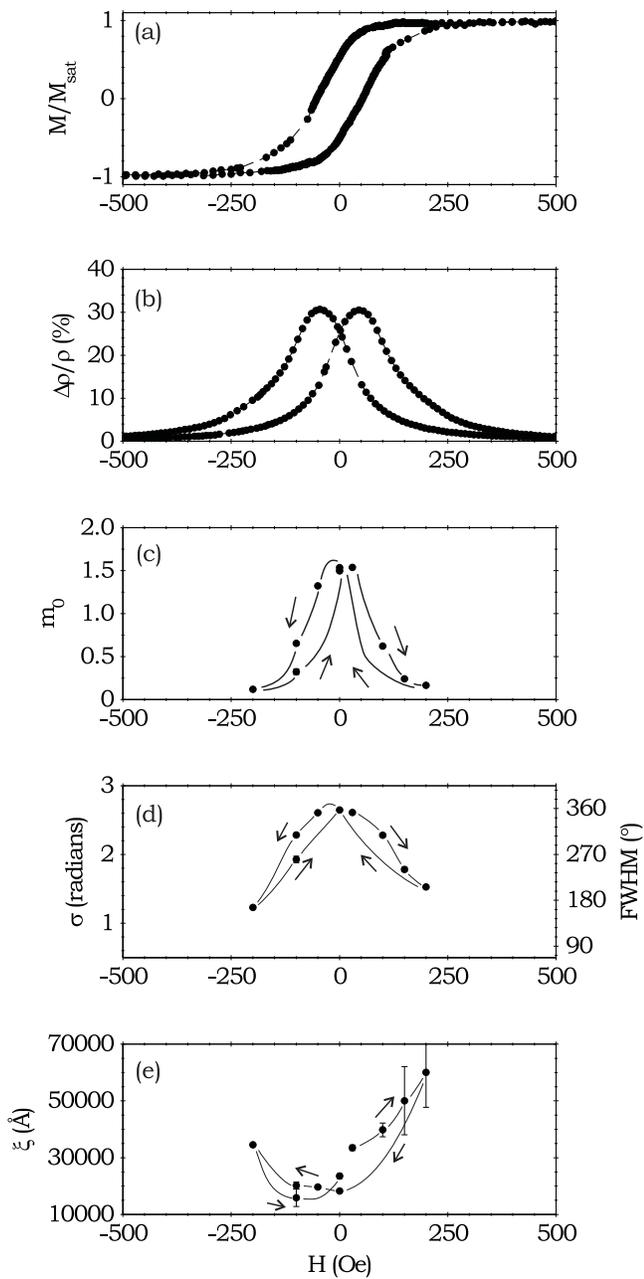,width=8.5cm} \caption{The diffuse
scattering observed at the AF peak as a function of applied field.
Each scan is offset for clarity.} \label{afpeak}
\end{figure}

A major conclusion of this paper results from the comparison of
the $Q_X$ distribution of the two peaks. The nuclear Bragg peak is
sharp but the AF peak is diffuse. The roughness is therefore
predominantly magnetic. We associate this magnetic roughness with
the existence of AF coupled domains. The diffuse scattering is
strongly peaked in $Q_Z$ therefore our data gives evidence for the
coherent coupling of the magnetic domains vertically through the
multilayer. Note, no evidence for diffuse scattering from
uncorrelated regions was observed which would be uniformly
distributed in $Q_Z$ \cite{savage}. Applying a saturating field
(Fig.\ref{frsm}(b)) destroys the AF correlations resulting in a
ferromagnetic alignment. Fig.\ref{afpeak} details sections in
$Q_X$ through the AF Bragg peak as one cycles from close to
remanence to positive saturation and then reverse to negative
saturation. At low fields ($<$100 Oe) the scattering is dominated
by the diffuse scattering. As the field is increased to saturation
only the specular ridge remains. Equivalent sections through the
nuclear Bragg peak reveal no evidence of diffuse scattering.

In order to quantitatively analyze our data and to characterize
the domain structure of the multilayer system, we now present a
new theoretical framework for  diffuse magnetic scattering
in systems with a spatially inhomogeneous magnetization profile,
${\bf m}({\bf r})$, where ${\bf r}$ is the position vector.
Considering a system where ${\bf m}({\bf r})$ is in-plane, we can
write $ {\bf m}({\bf r})=m_0 (\cos \phi({\bf r}),\sin \phi({\bf
r}) ,0)$, with phase angle, $\phi({\bf r})$, and amplitude, $m_0$.
Thus, we consider solely directional variations of ${\bf m}({\bf
r})$ which describe the different orientations of the magnetic
domains. We treat ${\bf m}({\bf r})$ and therefore $\phi({\bf r})$
as random variables, characterized by the correlation function
$C(|{\bf r}|)=\langle \phi({\bf r}) \phi(0) \rangle$. This is
similar to the treatment of structurally rough surfaces by Sinha
{\em et al.}\cite{sinha1}. Thus, $\phi({\bf r})$ plays a role
reminiscent to the local height variation in the non-magnetic case
and we parametrize $C(r)$ as:
\begin{equation}
 C(r)=\sigma^2 \exp\left(-r/\xi  \right) \, .
\end{equation}
$\sigma=\langle \phi^2\rangle$ is the width of the angular
distribution and therefore characterizes the {\em local magnetic
roughness}. $\xi$ is the lateral correlation length, i.e. a
measure for a typical domain size. We consider the magnetic
scattering function within the Born approximation, $S({\bf Q})
\propto \sum_{\alpha \beta} \int d^{3}{\bf r} e^{i {\bf Q}\cdot
{\bf r}} \left( \delta_{ \alpha \beta} -\hat{ Q}_\alpha \hat{
Q}_\beta \right) \langle m_\alpha({\bf r}) m_\beta(0)\rangle $,
where $ \hat{ Q}_\alpha$ is a unit vector component of the
transferred momentum, ${\bf Q}$. Performing the average with
respect to the different  domain orientations by assuming a
Gaussian distribution for $\phi({\bf r})$, we find in addition to
the specular scattering, $ S_{\rm spec.}({\bf Q})= m_0^2
e^{-\sigma^2}  \delta({\bf Q}_{\parallel})$, the diffusive
scattering function:
 \begin{eqnarray}
S_{\rm diff.}({\bf Q})&=& m_0^2 e^{-\sigma^2}
\int d^2 {\bf r} \, e^{i{\bf Q}_\parallel \cdot {\bf r}}
\left[  (1-\hat{Q}_X^2)  \,  {\rm sinh} \left(C(r)\right)
\right.
  \nonumber \\
& &
 \left.
 + (1-\hat{Q}_Y^2)  \,  2 {\rm sinh}^2 \left(C(r)/2\right)  \right] \,.
\label{diff_scatt}
\end{eqnarray}
Here, ${\bf Q}_\parallel$ is the in-plane component of ${\bf Q}$.
In our experimental geometry the detector aperture is set up such
that the neutron intensity is integrated out over $Q_Y$, which is
parallel to the applied field. Finally, when evaluated at the AF
ordering vector, it holds that $m_0=\mu \sin(\theta /2)$, where
$\mu$ is the Co magnetic moment and $\theta$ the angle between Co
moments in adjacent layers. Before  we  analyze our data using
Eq.~\ref{diff_scatt} we emphasize that, within the Gaussian
approximation, we treat the angle $\phi({\bf r})$, normally
restricted to $\pm \pi$, as an unrestricted variable. Therefore, we
cannot describe a system  with equally  distributed angles, i.e.
with $\langle e^{i \phi({\bf r})}\rangle=0$. However, this
practically  never occurs  after the system was exposed to an
external field, even if this field is set to zero or equal to the
coercive field, see also our results in Fig.~\ref{afpeak}, where
the AF peak always has some specular component. Furthermore, our 
result for the
diffusive magnetic scattering, Eq.~\ref{diff_scatt}, is
anisotropic with respect to $Q_x$ and $Q_y$. This is because
magnetic fluctuations perpendicular to the field are not only
larger in amplitude but also more extended in space compared to
those parallel to the field. In the present scattering geometry
this effect is negligible since $Q_x^2, Q_y^2 \ll Q_z^2\approx 1$.
However, for smaller $Q_z$ or by using polarized neutrons, we
predict a pronounced anisotropy of the diffuse magnetic scattering
if $S({\bf Q})$ is averaged with respect to  the component of
${\bf Q}$ parallel or perpendicular to the field.

\begin{figure}
\psfig{figure=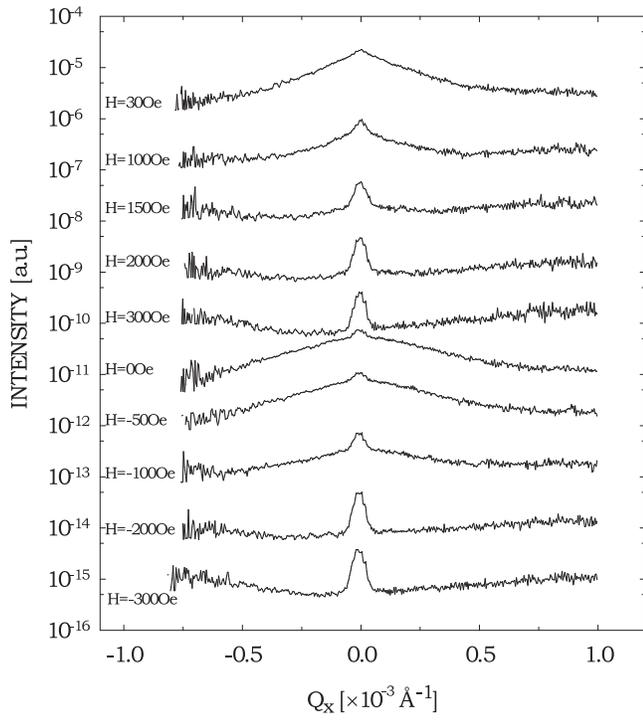,width=8.5cm}\caption{(a): The room
temperature MOKE magnetization loop for the
[Co(20\AA)/Cu(20\AA)]$\times$50 sample. (b) The magnetoresistance.
Panels (c,d,e) represent the parameters described in the text. The
lines are simply guides to the eye.} \label{mrmoke}
\end{figure}

The results of numerically convoluting the specular and diffuse
(Eq.\ref{diff_scatt}) contributions with the instrumental
resolution function and performing a least-squares fit to the data
are shown in Fig.~\ref{mrmoke}. The agreement between theory and experiment 
is excellent. Panels (a) and (b) display the
observed magnetization loop as measured by MOKE measurements and
the normalized change in resistivity respectively. The MOKE loop
and magnetoresistance curve both indicate good AF coupling. For
fields close to remanence the Co layers have a global
anti-parallel alignment(c) with a large magnetic roughness(d) and
a characteristic domain size of $\approx1\mu\mbox{m}$. For
increasing fields three effects occur. The anti-parallel alignment
across the non-magnetic spacer is diminished. The orientational
domain distribution within a given layer focuses around the
applied field direction and the domain size increases to
$\approx7\mu\mbox{m}$. Even at reasonably large fields the moments
are not perfectly aligned about the field direction. At +200 Oe
there still remains a substantial domain distribution although the
orientation of adjacent layers is nearly ferromagnetic ($m_0
\rightarrow 0$). At these fields the diffuse scattering approaches
the experimental background (primarily from incoherent scattering)
and represents the limits of the current measurements. For this
reason we cannot measure values of $\sigma$ close to zero in Fig.~
\ref{mrmoke} as saturation is approached. At saturation only the
structural specular peak remains. The data clearly show the
hysteresis in moving around the loop. The fact that all quantities
in Fig.~\ref{mrmoke} follow this hysteresis loop reveals the close
correlation between the GMR effect and the magnetic domain
correlations. The decrease of the angle $\theta$ between the
magnetization in neighboring layers causes not only the expected
decrease of the GMR but also a decreasing roughness accompanied by
an increase of the domain size. The latter effect is particularly
drastic for fields above the coercive field, $H_c$, i.e. $\xi
\approx \xi_0 f(H/H_c) $ with non-linearly growing function
$f(x)$. Below $H_c$ the dominant effect is the focusing of domain
orientations. The observed increase in domain size from $\sim$1.5
$\mu$m to 7 $\mu$m extracted from the data is typical of such
systems \cite{heyderman,herrmann}. Interestingly, qualitatively
similar effects have been observed around the
nuclear/ferromagnetic Bragg peak for an equivalent
ferromagnetically coupled system.

We now turn to the domain reversal mechanism. This can take place by 
directional magnetization fluctuations which are focused by the applied field.
or by domain wall motion where the
axis of magnetization is unchanged due to a dominating
magneto-crystalline anisotropy. In this latter scenario $\sigma$ would
be a measure for the probability of the magnetization aligning
anti-parallel to the external field rather than the width of a
directional distribution. However, our samples are only weakly
anisotropic. Imaging of 2nd AF coupled
multilayers\cite{aitchison} strongly supports a domain reversal
predominantly via rotation of the magnetization in agreement with
our analysis.

Our results can be compared with the recent work of Borchers {\it et
al.} (Ref.~\onlinecite{borchers}) on a weakly coupled system. Their 
data show that the reduction in the GMR from
the as-prepared state to the coercive state can be understood as a
loss of coherence of the AF coupling. The more strongly coupled
samples investigated here clearly show magnetically correlated
domains at the coercive field. By increasing the exchange coupling
by replacing Cu with Ru \cite{parkin2} there is a significant
change in the domain structure. Even at remanence the correlation
length is $\xi=(7 \pm 3)\times 10^4$ \AA, with the diffuse tail
extending across the whole of the $Q_X$ range. It is important to
note that only with the very long lateral coherence lengths of the
neutron beam used is it possible to measure $\xi$ accurately when
it is so large. A full discussion of the
dependence on spacer material and magnetic and spacer layer
thicknesses will be published separately \cite{langridge}.

To summarize, within a new theoretical framework we have
quantified the magnetic domain structure in an AF coupled
multilayer using diffuse magnetic neutron scattering. The
systematic study of the field dependence of the diffuse scattering
reveals a close relationship between magnetic roughness, domain
size, interlayer coupling and the GMR effect itself.

\acknowledgements The authors gratefully acknowledge fruitful
discussions with J.R.P. Webster, J. Penfold, S.W. Lovesey and S.K. Sinha. C.
H. Marrows would like to thank the Royal Commission for the
Exhibition of 1851 for financial support. We are grateful to the
Rutherford Appleton Laboratory for the provision of ISIS beamtime.


%
%

%
%

\end{document}